\documentclass[sigconf,nonacm]{acmart}

\usepackage{booktabs}
\usepackage{multirow}
\usepackage{url}
\usepackage{enumitem}   
\usepackage{makecell}   

\usepackage{bbding} 
\usepackage{epigraph}

\usepackage{colortbl} 
\definecolor{mygray}{rgb}{0.9, 0.9, 0.9}
\definecolor{myred}{rgb}{0.68627451, 0.14117647, 0.09803922}
\newcommand{\myeq}[1]{{Eq.~(\ref*{eq:#1})}}
\newcommand{\mysec}[1]{{Section~\ref*{sec:#1}}}
\newcommand{\mytable}[1]{{Table~\ref*{tab:#1}}}
\newcommand{\myfig}[1]{{Fig.~\ref*{fig:#1}}}

\AtBeginDocument{%
  }

\setcopyright{acmlicensed}
\copyrightyear{2018}
\acmYear{2018}
\acmDOI{XXXXXXX.XXXXXXX}
\acmConference[Conference acronym 'XX]{Make sure to enter the correct
  conference title from your rights confirmation email}{June 03--05,
  2018}{Woodstock, NY}
\acmISBN{978-1-4503-XXXX-X/2018/06}

\begin{document}

\title{Future-Conditioned Recommendations with  Multi-Objective Controllable Decision Transformer}








\author{Chongming Gao$^{1}$, Kexin Huang$^{1}$, Ziang Fei$^{1}$, Jiaju Chen$^{1}$, Jiawei Chen$^{2}$,\\Jianshan Sun$^{3}$, Shuchang Liu, Qingpeng Cai, Peng Jiang}

\affiliation{
  \institution{
  $^{1}$University of Science and Technology of China\\
  $^{2}$Zhejiang University\\
  $^{3}$Hefei University of Technology}
  \country{}
}


\renewcommand{\shortauthors}{Chongming Gao et al.}

\begin{abstract}
Securing long-term success is the ultimate aim of recommender systems, demanding strategies capable of foreseeing and shaping the impact of decisions on future user satisfaction. Current recommendation strategies grapple with two significant hurdles. Firstly, the future impacts of recommendation decisions remain obscured, rendering it impractical to evaluate them through direct optimization of immediate metrics. Secondly, conflicts often emerge between multiple objectives, like enhancing accuracy versus exploring diverse recommendations. Existing strategies, trapped in a ``training, evaluation, and retraining'' loop, grow more labor-intensive as objectives evolve.

To address these challenges, we introduce a future-conditioned strategy for multi-objective controllable recommendations, allowing for the direct specification of future objectives and empowering the model to generate item sequences that align with these goals autoregressively. We present the Multi-Objective Controllable Decision Transformer (MocDT), an offline Reinforcement Learning (RL) model capable of autonomously learning the mapping from multiple objectives to item sequences, leveraging extensive offline data. Consequently, it can produce recommendations tailored to any specified objectives during the inference stage. Our empirical findings emphasize the controllable recommendation strategy's ability to produce item sequences according to different objectives while maintaining performance that is competitive with current recommendation strategies across various objectives. 
\end{abstract}

\maketitle

\section{Introduction}

\setlength{\epigraphwidth}{.85\columnwidth}
\renewcommand{\epigraphflush}{center}
\renewcommand{\textflush}{flushepinormal}
\epigraph{\textit{``If we use, to achieve our purposes, a mechanical agency with whose operation we cannot efficiently interfere once we have started it, because the action is so fast and irrevocable that we have not the data to intervene before the actions complete, then we had better be quite sure that the purpose put into the machine is the purpose which we really desire and not merely a colorful imitation of it."} }
{{\footnotesize{\textit{––Norbert Wiener, In Science, 1960 \cite{wiener1960some}}}}}
	
\begin{figure}[!t]
\centering
\includegraphics[width=0.98\linewidth]{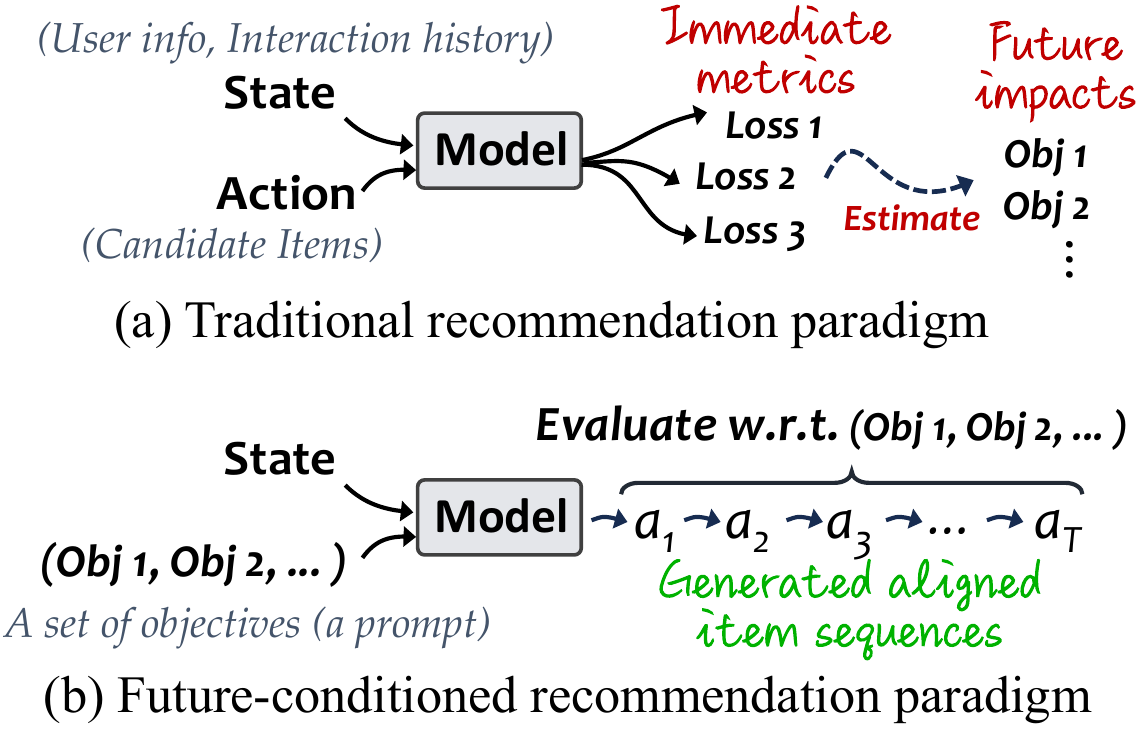}
\caption{Illustration of two paradigms in recommendation}
\label{fig:paradigms}
\end{figure}

Recommender systems facilitate seamless interaction between users and digital platforms, continuously adapting to reflect users' shifting preferences. Their primary objective is to curate a series of recommendations that not only meet immediate user needs but also lay the groundwork for ongoing platform engagement, thereby enhancing retention rates. However, the direct optimization of long-term engagement is fraught with challenges, as it involves predicting and measuring outcomes that lie in the future. As a workaround, these systems employ immediate feedback signals (e.g., clicks, purchases, and ratings) as surrogate measures to guide the optimization of future engagement.

This paradigm, while pragmatic, encounters notable challenges. Foremost is the task of quantifying the future impact of present actions, a hurdle that complicates the optimization of immediate performance metrics  \cite{surrogate, Ding:2023}. Furthermore, the imperative to reconcile multiple, often conflicting objectives (e.g., such as accuracy versus diversity) introduces further complexity. For instance, a strategy overly concentrated on enhancing immediate performance through continuous recommendations of previously preferred items may trigger user fatigue \cite{CIRS2023gao,gao2023alleviating}. This, in turn, can lead to a progressive decline in user engagement, underscoring the intricate balance required in recommendation strategies. 
Existing recommendation methods (shown in \myfig{paradigms}(a)), consequently, find themselves trapped in a continuous loop of recommendation and evaluation. As the number and complexity of objectives increase, this loop becomes increasingly burdensome, demanding more time and resources.

To address this challenge, we adopt a future-conditioned recommendation approach, as depicted in \myfig{paradigms}(b). This framework mandates that the model accepts a set of prioritized future objectives as input and generates item sequences that are aligned with these objectives. 
This methodology is inspired by the principles of upside-down Reinforcement Learning (RL), which focuses on mapping rewards to actions rather than forecasting the reward for each possible action. This philosophy represents a cornerstone of offline RL, aiming to derive a policy from historical interaction data rather than relying on online rewards \cite{offlineRLsurvey}. 

In recent years, many studies have formulated offline RL challenges as autoregressive sequence modeling problems using Decision Transformers (DT) \cite{NEURIPS2021_7f489f64} or Trajectory Transformers \cite{janner2021sequence}. Leveraging the formidable modeling and generalization capabilities of Transformer models, these approaches have reinterpreted reinforcement learning challenges through the lens of supervised learning methods. By doing so, they harness the predictive power inherent in Transformers to directly map sequences of states and actions to optimal outcomes, thereby circumventing the traditional complexities associated with RL problem-solving. 
Recent contributions by \citet{10.1145/3543507.3583418} and \citet{Wang2023CausalDT} have showcased the application of DT in crafting recommender systems. These efforts, however, are characterized by a limitation: they are tailored to scenarios with a single objective, thereby neglecting the complex multi-objective dynamics inherent to recommendation environments. Although some works \cite{liu2024sigir,10.1145/3627673.3679533} consider multiple objectives in their DT model, they did not focus on the requirement of changing objectives during the inference phase. Most importantly, current models focus on predicting the next item based on interaction contexts and specified rewards, leading to an evaluation framework centered on immediate performance metrics like NDCG and Recall. This narrow focus leaves the policy’s planning capabilities underexamined and fails to account for long-term objectives such as diversity, which are crucial in industrial recommender systems and require evaluation across a broader set of items.

We present the Multi-Objective Controllable Decision Transformer (MocDT) model, which innovatively applies a control signal for representing and prioritizing objectives. This allows the model to generate item sequences aligning with these goals in an autoregressive manner, eliminating the need for model retraining when the objectives change. The underlying concept draws a parallel to prompt-based techniques in large language models (LLMs), where prioritized objectives act as prompts to guide the model in planning and generating sequences of recommended items. By leveraging this approach, MocDT achieves a high degree of adaptability and controllability \cite{shen2024survey}.

It is well established that the efficacy of the DT model heavily depends on the quality of the offline dataset it is trained on \cite{bhargava2023should,when_rcsl_work}. In scenarios where recommendation datasets lack comprehensive offline interaction trajectories, we employ three data augmentation strategies to simulate diverse interaction patterns from existing offline data. These strategies are designed to enrich the dataset, enabling the model to better generalize across various recommendation objectives.

For evaluation, we rigorously test the model using a set of randomly selected objectives, requiring it to generate item sequences in a sequential manner, as illustrated in \myfig{paradigms}(b). To ensure a comprehensive analysis, we conduct experiments across five major categories of methods.
The success of MocDT is evaluated by its ability to produce trajectories that align with and fulfill the prioritized multiple objectives. This capability underscores the model's potential to serve as a highly controllable recommender system, which paves the way for more personalized and user-centric recommendation experiences.

The key contributions of this work are summarized as follows:
\begin{itemize}
    \item We introduce a novel future-conditioned recommendation framework that enables the inference-time generation of item sequences aligned with specified multiple objectives, thereby eliminating the need for model re-training when objectives evolve.
    \item We propose the Multi-Objective Controllable Decision Transformer (MocDT) approach that effectively maps a spectrum of objectives to the recommendation actions using offline interaction data.
    \item We develop three data augmentation strategies to enhance the diversity and quality of offline datasets, improving the overall effectiveness and robustness of the trained model.
    \item We conduct extensive empirical evaluations across five categories of methods, demonstrating the effectiveness of MocDT in generating trajectories that adhere to predefined multi-objective goals.
\end{itemize}

\section{Preliminary}

In this section, we provide an introduction to the fundamentals of offline reinforcement learning and the decision transformer.

\subsection{Offline Reinforcement Learning}
Reinforcement Learning (RL) problem is generally modeled via the Markov Decision Process (MDP): $\{\mathcal S, \mathcal A, T, r, \gamma\}$, where the $\mathcal{S}$ and $\mathcal{A}$ define the state and action space, $T(s_{t+1}|s_t,a_t)$ is the transition function, $r(s_t,a_t)$ serves as the reward function and $\gamma$ is the discount factor. 
The goal of RL is to derive a policy $\pi$ such that the expected cumulative reward $\mathbb E_{a_{t+1}\sim \pi(\cdot|s_t)}\sum_t\left[\gamma^t r(s_t,a_t)\right]$ is maximized. 
While classic RL relies on an iteration of policy learning and environment interaction, offline RL diverges by learning a policy from a fixed offline dataset $\mathcal D$. 
This dataset typically contains decision trajectories $\tau$ collected from some behavior policy $\pi_\beta$ that may not be fully known. 
Offline RL is particularly relevant in scenarios like healthcare and recommendation systems, where online interactions are either risky or expensive \cite{offlineRLsurvey}.

\subsection{Decision Transformer}
The Decision Transformer (DT) departs from the classical iterative policy-learning paradigm of RL. Instead, it reformulates the offline RL problem into a sequence modeling task. 
To elaborate, consider a trajectory in the offline dataset: 
$\tau = \langle s_1, a_1, g_1, s_2, a_2, g_2, \cdots\rangle,$
where $g_t = \sum_{t'=t}^{|\tau|}r_{t'}$ is the returns-to-go (RTG) from time $t$ of the trajectory $\tau.$ 
DT is a causal transformer network $\pi_\theta$ to predict $a_{t}$ conditioned by $s_t, g_t$, and the previous sequence $\tau_{1:t-1}.$
At training time, trajectories in the offline dataset are fed into the model, to minimize the empirical negative log-likelihood loss:
$$\mathcal L_\mathrm{DT} = -\mathbb E_{\tau\sim\mathcal D}\sum_{t=1}^{|\tau|} \log\pi_\theta(a_{t} | \tau_{1:t-1}, s_t, g_t).$$

\section{Method}
\label{sec:method}


In this work, we mainly focus on the multi-objective scenario of interactive recommendation, where a recommender policy iteratively suggests items to users and receives their feedback, continuing until the user session concludes.

In this paradigm, the user's historical interactions are modeled as the states $s_t$, and recommended items are treated as actions $a_t$, with some immediate performance indicators (e.g. CTR or a rating) being considered as the rewards $\boldsymbol{r}(s_t, a_t)$, this recommendation procedure is converted into a decision-making problem.

\begin{figure*}[!t]
    \centering
    \includegraphics[width=0.98\linewidth]{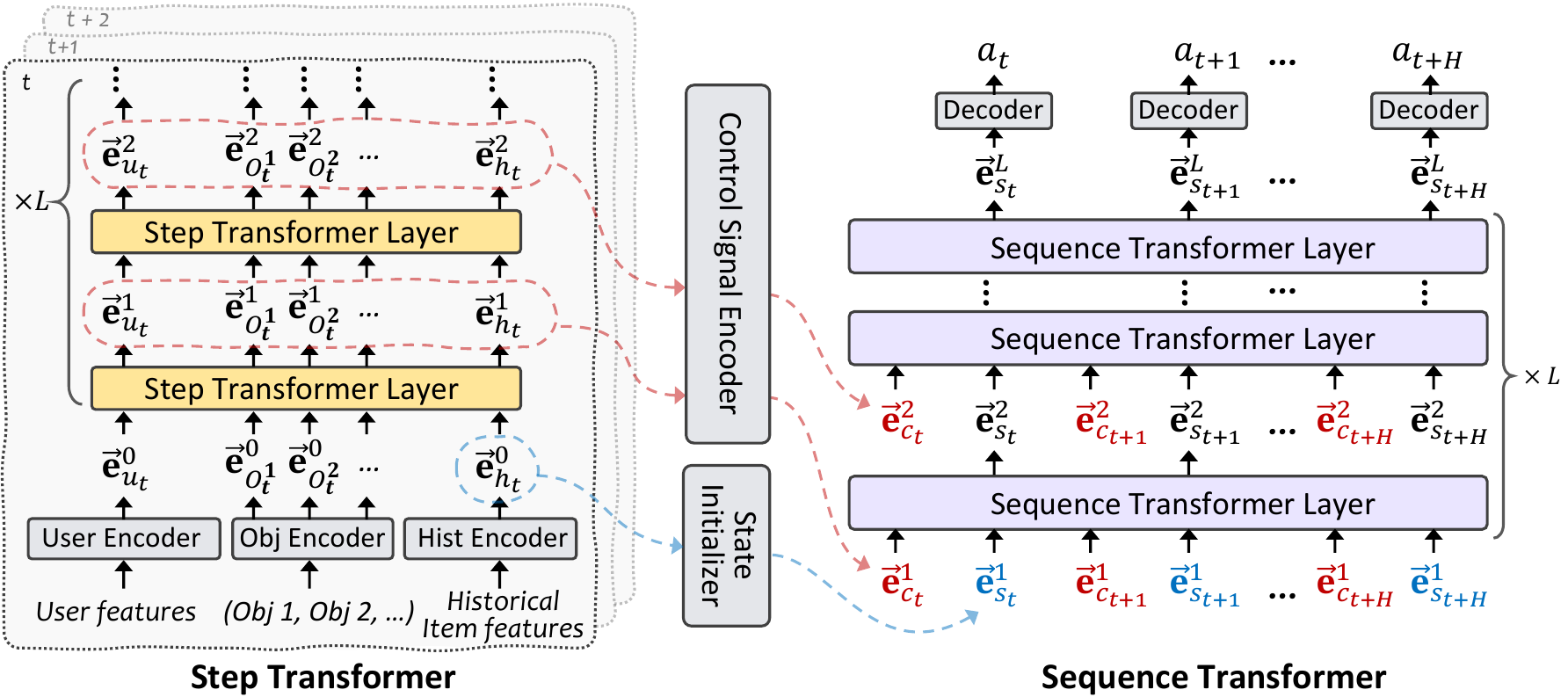}
    \caption{Architecture of the MocDT method.}
    \label{fig:framework}
\end{figure*}

Besides, recommendation settings might involve objectives (like diversity) that are evaluated at an aggregate level---across the span of the interaction trajectory, rather than on individual state-action pairs. Consequently, an immediate reward signal $r(s_t, a_t)$ can not reflect the feature of a whole trajectory $\tau$. 
To address these nuances, we introduce a future-conditioned recommendation framework and propose an adaption of the Decision Transformer (DT) tailored for recommendation contexts.

\subsection{DT as Future-Conditioned Recommender}
The key idea is to recommend conditioned on the future recommendation trajectory, considering both the cumulative individual-level rewards and the aggregate-level objectives that pertain to the to-go trajectory. Instead of measuring the whole to-go sequence like RTG, we employ a sliding window of the future trajectory, which is continuously evolving during the recommendation process. In this way, we do not restrict the policy to end in some pre-set rounds, hence mirroring the fluid nature of real-world recommender systems more closely.

Formally, denote an $H$-step future sequence after timestamp $t$ as $\tau_{t+1}^H = \langle \tau_{t+1}, \dots, \tau_{t+H} \rangle$, 
upon which several recommendation objectives $(O_1(\tau_{t+1}^H), O_2(\tau_{t+1}^H), \cdots)$ are measured.
These objectives may rely on diverse functions for computation. 
We introduce two examples of two commonly recognized objectives in recommendation settings: cumulative rating and diversity.

\smallskip \noindent $\bullet$ \textit{Cumulative Rating.}
Assuming $r(s_t, a_t)$ represents the user's rating\footnote{Here, we collectively refer to other accuracy-like metrics, such as impression and click-through rate (CTR), as ratings.} on item $a_t$, the cumulative rating is:
\begin{equation}\label{eq:def_rating}
    O_\text{rate}(\tau_{t+1}^H) = \sum_{t'=t+1}^{t+H} r(s_{t'}, a_{t'}).
\end{equation}
\smallskip 
\noindent $\bullet$ \textit{Diversity.}
Using $J(A, B)$ to denote the Jaccard similarity over the categories of two item sets $A, B$, then the average diversity can be quantified by:
\begin{equation}\label{eq:def_diversity}
    O_{\text{div}}(\tau_{t+1}^H) = \frac1H\sum_{t'=t+1}^{t+H}\left(1 - J(\{a_i\}_{i=t+1}^{t'-1}, \{a_{t'}\})\right).
\end{equation}

\smallskip
The recommender policy aims to generate actions steering towards a future trajectory aligned with our goals. Given the inherent conflict among multiple objectives, achieving an optimal state for all is impractical. Thus, we prioritize certain objectives and define a control signal $c_t$, which is then used to guide the model to generate recommendations that align with these priorities.

\begin{definition}[Control Signal]
    The control signal $c_t$ is a representation of the anticipated future trajectory, constructed from the current interaction state $s_t$ and a set of anticipated objective values $(O_1(\tau_{t+1}^H), O_2(\tau_{t+1}^H), \cdots)$. It is parameterized by the model $g_\phi$:
    \begin{equation}\label{eq:future_ctrl}
        c_t = g_\phi(s_t, \langle O_1(\tau_{t+1}^H), O_2(\tau_{t+1}^H), \cdots \rangle).    
    \end{equation}
\end{definition}

Similar to existing DT methods, we utilize the negative log-likelihood loss to train a Transformer-based policy network $\pi_\theta$. The loss function is articulated as:
\begin{equation}\label{eq:loss_fn}
    \mathcal L_\mathrm{MocDT} = -\mathbb E_{\tau\sim\mathcal D}\sum_{t=1}^{|\tau|-H}\log\pi_\theta(a_t|\tau_{1:t-1}, s_t, c_t).
\end{equation}

This formulation leverages the multi-objective control signal $c_t$, enabling the policy $\pi_\theta$ to generate actions that not only resonate with the prioritized objectives but also ensure the recommender system's adaptability.

\subsection{Model Architecture}


The MocDT model's architecture, as illustrated in \myfig{framework}, is comprised of the following critical components, discussed in the order of the data flow:
\begin{itemize}[leftmargin=*]
\item \textit{Feature Encoder} encompasses three sub-modules: the \textit{User Encoder}, \textit{Objective Encoder}, and \textit{History Encoder}. \textit{User Encoder} transforms the raw user information at step $t$, including the User ID and various available user features, into a user embedding vector $\vec{\mathbf{e}}^0_{{u}_{t}}$. The \textit{Objective Encoder} encodes the normalized values of multiple objectives $(O^1_t, O^2_t, \cdots)$ of step $t$ into corresponding objective embedding vectors $\langle\vec{\mathbf{e}}^0_{{O}^1_{t}}, \vec{\mathbf{e}}^0_{{O}^2_{t}}, \cdots\rangle$, with all input objective values normalized between [0, 1].
The \textit{History Encoder}, utilizing a GRU model, transcodes the sequence of user-item interactions into an initial history embedding vector $\vec{\mathbf{e}}^0_{{h}_{t}}$.
\item \textit{Step Transformer} processes the encoded embedding vectors $(\vec{\mathbf{e}}^0_{{u}_{t}}, \allowbreak \langle\vec{\mathbf{e}}^0_{{O}^1_{t}}, \vec{\mathbf{e}}^0_{{O}^2_{t}}, \cdots\rangle, \vec{\mathbf{e}}^0_{{h}_{t}})$ by passing them through $L$ Transformer layers to enrich their representation. The output is then funneled into the \textit{Control Signal Encoder} to construct the control signal $c_t$ defined in \myeq{future_ctrl}.
\item \textit{{Control Signal Encoder}} operationalizes $g_\phi$ from \myeq{future_ctrl}. Here, we simply implement it by concatenating the representation vectors $(\vec{\mathbf{e}}^l_{{u}_{t}}, \langle\vec{\mathbf{e}}^l_{{O}^1_{t}}, \vec{\mathbf{e}}^l_{{O}^2_{t}}, \cdots\rangle, \vec{\mathbf{e}}^l_{{h}_{t}})$ of the $l$-th layer of the \textit{Step Transformer} and passing them through an MLP to obtain the initial control representation vector $\vec{\mathbf{e}}^l_{{c}_{t}}$ for the \textit{Sequence Transformer}.
\item \textit{{State Initializer}} further refines the history embedding vector $\vec{\mathbf{e}}^0_{{h}_{t}}$ into the initial state vector $\vec{\mathbf{e}}^1_{{s}_{t}}$ for the \textit{Sequence Transformer}. 
\item \textit{{Sequence Transformer}} is the keystone of our MocDT model, integrating initial inputs from the \textit{Control Signal Encoder} and the \textit{State Initializer}. It systematically generates item sequences by propagating through its $L$ Transformer layers, culminating in a horizon of $t+H$ steps. Each iterative step is decoded through a \textit{Decoder} to yield a specific action (i.e., recommending an item), aligning the sequence with pre-set objectives $(O^1_t, O^2_t, \cdots)$.
\end{itemize}

\smallskip \noindent \textbf{Remark.} Typically, the trajectory in DT is composed of the state-action-return sequence $\langle s_t, a_t, g_t \rangle.$ Except for replacing RTG $g_t$ with the control signal $c_t$, we further reduce the action (i.e. recommended item) from the trajectory, since it has already been modeled in the state by historical items.



\subsection{Enhance MocDT with Synthetic Data}
It has been highlighted in \citet{when_rcsl_work} that the efficacy of a trained DT model relies on the quality of the offline dataset it is trained on. The model's ability to derive optimal policies is compromised if the training data lacks a comprehensive representation of optimal decision trajectories.

Intuitively, the loss function defined in Eq~\eqref{eq:loss_fn} aims to align the model's conditioned prediction with the probabilities induced by the behavior policy $\mathrm{P}_\beta(a_t|\tau_{1:t-1}, s_t, c_t)$, which can be re-formulated via Bayes's rule as (subscripts are omitted for simplicity):
\begin{equation}\label{eq:dt_reformulate}
    \mathrm{P}_\beta(a|\tau, s, c) = \frac{\mathrm{P}_\beta(a, \tau, s) \mathrm{P}_\beta(c | a, \tau, s)}{\mathrm{P}_\beta(\tau, s) \mathrm{P}_\beta(c|\tau, s)}
= \pi_\beta(a|s) \frac{\mathrm{P}_\beta(c | a, \tau, s)}{\mathrm{P}_\beta(c|\tau, s)},
\end{equation}
where the last equation holds for the Markov property. 
Eq~\eqref{eq:dt_reformulate} indicates that the learned policy $\pi_\theta$ reweights the behavior policy $\pi_\beta$, favoring actions $a_t$ that are more likely to fulfill the specified future condition $c_t$. Thus it's natural that the learned policy may not operate well in the absence of comprehensive offline trajectories, since lack of an action $a_t$ to guide the conditioned policy.

To mitigate this issue, one plausible solution is to utilize synthetic data as a means of augmenting the offline dataset $\mathcal D$, which has demonstrated effectiveness in previous works \citep{tra_stiching_1, tra_stiching_2}. Introducing synthetic sequences that mirror desired future trajectories invites the learned policy to select suitable actions, filling the gaps left by the original dataset's limitations.
In this work, we mainly consider the combination of cumulative rating and category diversity defined in Eq~\eqref{eq:def_rating} and Eq~\eqref{eq:def_diversity} for the multi-objective setting. These dual objectives encapsulate the frequent balancing challenge faced by recommendation systems \citep{jannach2023survey}.
To generate synthetic data that reflects these differing objectives, we have developed three data augmentation strategies. 


Consider a user $u$'s recommendation trajectory $\tau$ in the offline dataset $\mathcal D$, At each timestamp $t$, the user has already been recommended with a list of history items $h_t = \{a_{t'}|t'\le t-1\}$. We construct a synthetic to-go sequence from this timestamp $t$, denoted as $\hat \tau_\mathrm{togo}$, which is then concatenated with the original trajectory: $\langle \tau_{1:t-1}, \hat \tau_\mathrm{togo} \rangle$ as the augmented trajectory.
Every item presented in the to-go sequence $\hat \tau_\mathrm{togo}$ is selected from the user's interacted items record in offline dataset $\mathcal D$ and the to-go sequence is consistently set to a length of $2H$, with tailored selection criteria for each strategic goal:
\begin{itemize}[leftmargin=*]
    \item For \textbf{cumulative rating}, the to-go sequences are constructed to optimize this objective. At each subsequent timestamp $t' \ge t$, we select the next action as the item with the highest rating not yet featured in the evolving recommended item list $h_{t'}$.
    \item For \textbf{diversity}, we adopt a greedy strategy that strives to diversify the list of presented categories in the trajectory. At every timestamp $t' \ge t$, an item with the least category overlap with recommended items $h_{t'}$ is picked as the next action. If all categories are represented in $h_{t'}$, we simply reset this list as $h_{t'} = \emptyset$, and the selection process starts anew, disregarding previous choices. 
    \item Additionally, a \textbf{random} strategy is implemented, where items are indiscriminately selected and incorporated into the to-go sequence. This random approach ensures an unbiased enhancement of the offline data.
\end{itemize}
In essence, these three augmentation strategies employ pre-existing interactions from the dataset to craft alternative future recommendation paths for each state $s_t$. Empirically, this method facilitates some kind of exploration by synthetic data within the boundaries of the available offline dataset.

Contrary to previous methodologies \citep{tra_stiching_1, tra_stiching_2} that typically leveraged a trained model to fabricate additional data, our approach harnesses direct interactions from the existing dataset. This strategy is not only more straightforward but also avoids the complexities associated with additional modeling steps.

\begin{table}[t]
\caption{Statistics of three datasets.}
\renewcommand\arraystretch{1.3}
\label{tab:dataset}
\tabcolsep=2pt
\begin{tabular}{@{}ccccc@{}}
\toprule
\textbf{Datasets}                   & \textbf{\#Users} & \textbf{\#Items} & \textbf{\#Interactions} & \textbf{\#Categories} \\ \midrule
{MovieLens-1M}   & 6,040   & 3,706  & 1,000,209     & 19           \\
{KuaiRand-Pure}   & 7,176   & 10,728  & 12,530,806     & 31           \\
{Zhihu-1M}  & 27,285  & 7,551   & 1,436,609      & 46    \\       
\bottomrule
\end{tabular}
\end{table}

\section{Experiments}
\label{sec:exp}

We are interested in the following research questions:

\begin{itemize}[leftmargin=*]
    \item \textbf{(RQ1)} How does MocDT perform compared to state-of-the-art recommender methods?
    \item \textbf{(RQ2)} How effectively can the MocDT model control the desired long-term goals by adhering to specific objectives?
    \item \textbf{(RQ3)} What are the effects of the augmentation strategy and augmentation rate?
    \item \textbf{(RQ4)} What are the effects of the key parameters?
\end{itemize}

Finally, we present a case study to illustrate how the model operates in practice.



\begin{table*}[t]
\caption{Performance on three datasets in Diversity-focused and Cumulative Rating-focused scenarios.}
\label{tab:data}
\tabcolsep=2pt
\renewcommand\arraystretch{1.2}
\begin{tabular}{lcccccccccccc}
    \toprule
    \multirow{3}{*}{\textbf{Method}} & \multicolumn{4}{c}{\textbf{MovieLens-1M}}                                                                               & \multicolumn{4}{c}{\textbf{KuaiRand-Pure}}                                                                              & \multicolumn{4}{c}{\textbf{Zhihu-1M}}                                                                                            \\ \cmidrule(lr){2-5} \cmidrule(lr){6-9} \cmidrule(lr){10-13}
                           & \multicolumn{2}{c}{Focus on Diversity}                      & \multicolumn{2}{c}{Focus on Rating}                         & \multicolumn{2}{c}{Focus on Diversity}                      & \multicolumn{2}{c}{Focus on Rating}                         & \multicolumn{2}{c}{Focus on Diversity}                      & \multicolumn{2}{c}{Focus on Rating}                         \\ \cmidrule(lr){2-3} \cmidrule(lr){4-5} \cmidrule(lr){6-7} \cmidrule(lr){8-9} \cmidrule(lr){10-11} \cmidrule(lr){12-13}
                           & \multicolumn{1}{c}{Diversity} & \multicolumn{1}{c}{\scriptsize{Rating}} & \multicolumn{1}{c}{\scriptsize{Diversity}} & \multicolumn{1}{c}{Rating} & \multicolumn{1}{c}{Diversity} & \multicolumn{1}{c}{\scriptsize{Rating}} & \multicolumn{1}{c}{\scriptsize{Diversity}} & \multicolumn{1}{c}{Rating} & \multicolumn{1}{c}{Diversity} & \multicolumn{1}{c}{\scriptsize{Rating}} & \multicolumn{1}{c}{\scriptsize{Diversity}} & \multicolumn{1}{c}{Rating} \\ \toprule
    \rowcolor{mygray}   
    \multicolumn{5}{l}{\textbf{Sequential Recommender}} &&&&&&&&\\                     
    BERT4Rec                & 0.80\tiny{$\pm$0.00}                   & 28.86\tiny{$\pm$5.30}               & 0.80\tiny{$\pm$0.00}                   & 28.86\tiny{$\pm$5.30}               & 0.94\tiny{$\pm$0.00}                   & 4.11\tiny{$\pm$1.78}                & 0.94\tiny{$\pm$0.00}                   & 4.11\tiny{$\pm$1.78}                & \textbf{1.00\tiny{$\pm$0.00}}          & 4.40\tiny{$\pm$1.47}                & \textbf{1.00\tiny{$\pm$0.00}}          & 4.40\tiny{$\pm$1.47}                \\ \toprule
    \rowcolor{mygray}
    \multicolumn{5}{l}{\textbf{Multi-objective Recommender}} &&&&&&&&\\                     
    SingleTask              & 0.64\tiny{$\pm$0.05}                   & 40.94\tiny{$\pm$5.87}               & 0.72\tiny{$\pm$0.04}                   & 40.50\tiny{$\pm$5.95}               & 0.80\tiny{$\pm$0.12}                   & 5.74\tiny{$\pm$1.88}                & 0.87\tiny{$\pm$0.00}                   & 5.38\tiny{$\pm$1.90}                & \textbf{1.00\tiny{$\pm$0.00}}          & 4.25\tiny{$\pm$1.52}                & 1.00\tiny{$\pm$0.00}                   & 4.02\tiny{$\pm$1.58}                \\
    SharedBottom            & 0.67\tiny{$\pm$0.05}                   & 40.43\tiny{$\pm$5.90}               & 0.67\tiny{$\pm$0.05}                   & 40.43\tiny{$\pm$5.90}               & 0.88\tiny{$\pm$0.00}                   & 3.31\tiny{$\pm$1.71}                & 0.94\tiny{$\pm$0.06}                   & 5.41\tiny{$\pm$1.83}                & \textbf{1.00\tiny{$\pm$0.00}}          & 3.98\tiny{$\pm$1.60}                & 0.97\tiny{$\pm$0.00}                   & 4.31\tiny{$\pm$1.66}                \\
    ESMM                    & 0.71\tiny{$\pm$0.00}                   & 41.08\tiny{$\pm$5.56}               & 0.65\tiny{$\pm$0.03}                   & 40.51\tiny{$\pm$6.02}               & 0.91\tiny{$\pm$0.00}                   & 5.40\tiny{$\pm$1.90}                & 0.91\tiny{$\pm$0.00}                   & 5.40\tiny{$\pm$1.90}                & \textbf{1.00\tiny{$\pm$0.00}}          & 4.76\tiny{$\pm$1.66}                & \textbf{1.00\tiny{$\pm$0.00}}          & 4.46\tiny{$\pm$1.56}                \\
    MMOE                    & 0.67\tiny{$\pm$0.06}                   & 40.53\tiny{$\pm$5.86}               & 0.67\tiny{$\pm$0.06}                   & 40.53\tiny{$\pm$5.86}               & 0.88\tiny{$\pm$0.08}                   & \textbf{5.80\tiny{$\pm$1.75}}       & 0.94\tiny{$\pm$0.00}                   & 5.09\tiny{$\pm$1.87}                & 0.98\tiny{$\pm$0.00}                   & 4.08\tiny{$\pm$1.56}                & 0.98\tiny{$\pm$0.00}                   & 4.15\tiny{$\pm$1.56}                \\
    PLE                     & 0.63\tiny{$\pm$0.05}                   & 40.66\tiny{$\pm$5.66}               & 0.67\tiny{$\pm$0.05}                   & \textbf{40.58\tiny{$\pm$5.82}}      & 0.94\tiny{$\pm$0.02}                   & 4.37\tiny{$\pm$1.90}                & 0.91\tiny{$\pm$0.00}                   & 5.43\tiny{$\pm$1.88}                & 0.99\tiny{$\pm$0.00}                   & 4.09\tiny{$\pm$1.58}                & 0.99\tiny{$\pm$0.00}                   & 4.09\tiny{$\pm$1.58}                \\ \toprule
    \rowcolor{mygray}
    \multicolumn{5}{l}{\textbf{Multi-objective RL-based Recommender}} &&&&&&&&\\                     
    RMTL\_ESMM              & 0.69\tiny{$\pm$0.04}                   & 39.10\tiny{$\pm$5.66}               & 0.66\tiny{$\pm$0.05}                   & 39.42\tiny{$\pm$5.81}               & 0.93\tiny{$\pm$0.01}                   & 5.18\tiny{$\pm$1.95}                & 0.95\tiny{$\pm$0.06}                   & \textbf{5.53\tiny{$\pm$1.83}}       & 0.99\tiny{$\pm$0.00}                   & 4.69\tiny{$\pm$1.67}                & 0.97\tiny{$\pm$0.00}                   & 5.08\tiny{$\pm$1.68}                \\
    RMTL\_MMOE              & 0.70\tiny{$\pm$0.05}                   & 40.42\tiny{$\pm$5.75}               & 0.69\tiny{$\pm$0.09}                   & 40.32\tiny{$\pm$5.98}               & 0.93\tiny{$\pm$0.05}                   & 5.49\tiny{$\pm$1.86}                & 0.95\tiny{$\pm$0.04}                   & 5.22\tiny{$\pm$1.93}                & \textbf{1.00\tiny{$\pm$0.00}}          & 3.97\tiny{$\pm$1.43}                & \textbf{1.00\tiny{$\pm$0.00}}          & 4.78\tiny{$\pm$1.61}                \\
    RMTL\_PLE               & 0.75\tiny{$\pm$0.01}                   & \textbf{41.20\tiny{$\pm$5.49}}      & 0.72\tiny{$\pm$0.12}                   & 39.40\tiny{$\pm$6.54}               & 0.93\tiny{$\pm$0.04}                   & 5.18\tiny{$\pm$1.80}                & \textbf{0.95\tiny{$\pm$0.03}}          & 5.17\tiny{$\pm$1.87}                & \textbf{1.00\tiny{$\pm$0.00}}          & 3.80\tiny{$\pm$1.51}                & 1.00\tiny{$\pm$0.00}                   & 4.21\tiny{$\pm$1.54}                \\ \toprule
    \rowcolor{mygray}
    \multicolumn{5}{l}{\textbf{Multi-objective Controllable Recommender}} &&&&&&&&\\                     
    PHN\_LS                 & 0.83\tiny{$\pm$0.04}                   & 35.25\tiny{$\pm$4.77}               & 0.84\tiny{$\pm$0.04}                   & 35.33\tiny{$\pm$4.70}               & 0.92\tiny{$\pm$0.05}                   & 4.88\tiny{$\pm$1.87}                & 0.92\tiny{$\pm$0.05}                   & 4.88\tiny{$\pm$1.86}                & \textbf{1.00\tiny{$\pm$0.00}}          & 4.80\tiny{$\pm$1.67}                & \textbf{1.00\tiny{$\pm$0.00}}          & 4.84\tiny{$\pm$1.67}                \\
    PHN\_EPO                & 0.84\tiny{$\pm$0.06}                   & 34.93\tiny{$\pm$4.66}               & \textbf{0.85\tiny{$\pm$0.06}}          & 35.01\tiny{$\pm$4.64}               & 0.92\tiny{$\pm$0.04}                   & 4.74\tiny{$\pm$1.85}                & 0.93\tiny{$\pm$0.04}                   & 4.89\tiny{$\pm$1.86}                & 1.00\tiny{$\pm$0.01}                   & \textbf{4.95\tiny{$\pm$1.66}}       & 1.00\tiny{$\pm$0.01}                   & 4.91\tiny{$\pm$1.66}                \\ \toprule
    \rowcolor{mygray}
    \multicolumn{5}{l}{\textbf{DT-based Recommender}} &&&&&&&&\\
    DT4Rec                  & 0.86\tiny{$\pm$0.24}                   & 31.16\tiny{$\pm$8.67}               & 0.33\tiny{$\pm$0.25}                   & 31.85\tiny{$\pm$7.85}               & 0.94\tiny{$\pm$0.09}                   & 4.51\tiny{$\pm$1.91}                & 0.88\tiny{$\pm$0.09}                   & 4.56\tiny{$\pm$1.90}                & 0.98\tiny{$\pm$0.03}                   & 4.70\tiny{$\pm$1.70}                & 0.98\tiny{$\pm$0.02}                   & 5.09\tiny{$\pm$1.69}                \\
    CDT4Rec                 & 0.71\tiny{$\pm$0.13}                   & 31.61\tiny{$\pm$6.99}               & 0.70\tiny{$\pm$0.13}                   & 31.56\tiny{$\pm$7.03}               & 0.88\tiny{$\pm$0.06}                   & 4.42\tiny{$\pm$1.79}                & 0.86\tiny{$\pm$0.08}                   & 4.39\tiny{$\pm$1.86}                & 0.99\tiny{$\pm$0.01}                   & 4.76\tiny{$\pm$1.66}                & 0.98\tiny{$\pm$0.03}                   & \textbf{5.14\tiny{$\pm$1.72}}       \\ \toprule
    \rowcolor{mygray}
    \multicolumn{6}{l}{\textbf{Multi-objective Controllable DT-based Recommender}} &&&&&&&\\
    MocDT (Ours)                           & \textbf{0.87\tiny{$\pm$0.05}}          & 25.56\tiny{$\pm$5.58}               & 0.72\tiny{$\pm$0.23}                   & 40.57\tiny{$\pm$4.73}               & \textbf{0.95\tiny{$\pm$0.04}}          & 4.22\tiny{$\pm$1.90}                & 0.38\tiny{$\pm$0.23}                   & 5.42\tiny{$\pm$1.95}                & \textbf{1.00\tiny{$\pm$0.00}}          & 4.23\tiny{$\pm$1.76}                & 0.98\tiny{$\pm$0.02}                   & 5.10\tiny{$\pm$1.68}                \\ \cmidrule(lr){2-13}
    Ranking (Ours)                         & \textbf{1}                             & \scriptsize{14}                         & \scriptsize{5}                             & \textbf{2}                          & \textbf{1}                             & \scriptsize{12}                         & \scriptsize{14}                            & \textbf{3}                          & \textbf{1}                             & \scriptsize{9}                          & \scriptsize{10}                            & \textbf{2}                          \\ \bottomrule
    \end{tabular}
\end{table*}

\subsection{Experimental Setup}

\subsubsection{Metrics}

Accuracy and diversity are widely acknowledged as conflicting metrics within recommender systems. To assess these dimensions, we compute the listwise cumulative rating and diversity metrics, as formalized in \myeq{def_rating} and (\ref{eq:def_diversity}).

\subsubsection{Datasets}
\label{seq:env}




In our study, we leverage three public recommendation datasets---MovieLens-1m, KuaiRand-Pure \cite{gao2022kuairand}, and Zhihu-1m \cite{hao2021largescale}---whose fundamental statistics are encapsulated in \mytable{dataset}. Each dataset is inclusive of item categories, which are essential for computing the item distance integral to the diversity metric. 

A notable challenge in offline recommendation systems is the inherent sparsity of data, which obscures the evaluation of unrecorded user-item interactions. To mitigate this, we enrich the datasets to construct fully-observed user-item interaction matrices prior to evaluation \cite{jinhuang22,Keeping-recsys}, employing matrix factorization for MovieLens and the Wide \& Deep model \cite{Cheng2016WideD} for KuaiRand and Zhihu. The predictive algorithms are finely tuned to minimize mean absolute error on the validation set, enhancing the reliability of the completed data, as detailed in our supplementary code. It is noteworthy that the augmented data is solely utilized for evaluating the MocDT and baseline models, without influencing their training phases.

\subsubsection{Baselines}
We select one classic sequential recommender, five multi-objective recommenders, one multi-objective RL-based recommender, two multi-objective controllable algorithms, and two DT-based recommenders as baselines, which are introduced as follows. All implementation details can be found via our repository\footnote{\url{https://anonymous.4open.science/r/MocDT}}.

\begin{itemize}[leftmargin=*]
    \item \textit{Sequential Recommender}: 
    \textbf{BERT4Rec} \cite{Sun2019BERT4RecSR}, a classical sequential recommender that employs a deep bidirectional self-attention mechanism to model user behavior sequences.
    \item \textit{Multi-Objective Methods}: 
    \textbf{Single Task} \cite{luong2015multi} is a naive multi-task method that learns each task separately.
    \textbf{Shared Bottom} \cite{Ma2018ModelingTR} is a basic multi-task model with a shared bottom structure.
    \textbf{ESMM} \cite{Ma2018EntireSM} is a multi-task model that uses auxiliary tasks to help learn the main task.
    \textbf{MMoE} \cite{Ma2018ModelingTR}, a multi-gate mixture-of-experts framework that learns multiple tasks simultaneously by sharing the bottom sub-model across all tasks while having a gating network for each task.
    \textbf{PLE} \cite{Tang2020ProgressiveLE} is a multi-task model that separates shared components and
task-specific components and adopts a progressive routing mechanism to extract knowledge.
    \item \textit{Multi-Objective RL-based Recommender}: 
    \textbf{RMTL} \cite{LiuRLMultiTaskRec} is an RL-enhanced framework that fine-tunes existing multi-task methods by leveraging the weights provided by critic networks.
    \item \textit{Controllable Methods}: 
        \textbf{PHN} \cite{Navon2020LearningTP} is a multi-objective approach that learns the entire Pareto front simultaneously using a single hypernetwork. which receives a desired preference as an input vector and returns a Pareto-optimal model. \textbf{PHN\_LS} and \textbf{PHN\_EPO} are two variants using linear scalarization and exact Pareto optimal \cite{pmlr-v119-mahapatra20a}, respectively.
    \item \textit{Decision Transformer-based Recommenders}: 
    \textbf{DT4Rec} \cite{Zhao2023UserRR} employs DT with a focus on optimizing user retention, by designing a reward embedding aggregation module and utilizing weighted contrastive learning.
    \textbf{CDT4Rec} \cite{Wang2023CausalDT}, employs DT in recommendation with a focus on capturing the causal logic behind users' observed behaviors. Besides predicting the next action, CDT4Rec also predicts the next immediate reward.
\end{itemize}

\subsection{Overall Performance Comparison (RQ1)}

Our evaluation encompassed all proposed methods across three datasets, as presented in \mytable{data}. For each method, performance was gauged by analyzing the entire test user base, computing mean and standard deviation for outcomes associated with the generated sequences of $H=10$ items. This evaluation distinctly categorizes methods based on their ability to modulate outcomes in terms of Cumulative Rating (abbreviated as ``Rating'') and Diversity. 
We only describe the setting in Diversity-focused scenarios since it is the same way in Rating-focused scenarios.

Unlike PHN and MocDT, other models do not possess the ability to adjust these metrics directly. For such models, an average Diversity score was computed from a designated experimental group (with different seeds and initialization), and a corresponding result with a similar Diversity score and its Rating was selected. Note each (Rating, Diversity) pair is from the same run.
With PHN, we aimed to calculate the max Diversity across 5 specified objective points $(O_{\text{rating}}, O_{\text{div}})\in \{(0.0, 1.0), (0.1, 0.9), (0.2, 0.8), (0.3, 0.7), (0.4, 0.6)\}$, determining corresponding Ratings via a similar methodology to other methods. 
For MocDT, we found the maximal Diversity objective values under objective points $\{(0.0, 1.0), (0.5, 1.0), (1.0, 1.0)\}$ and then located the corresponding Rating values.

The results show that MocDT excels in two key aspects: multi-objective consideration and explicit control over multiple objectives. Unlike classical sequential models such as Bert4Rec, DT4Rec, and CDT4Rec, which focus on optimizing a single objective like cumulative ratings, MocDT is designed to handle multiple objectives simultaneously. This gives it a distinct advantage in complex scenarios where balancing different objectives, such as Rating and Diversity, is crucial.

Additionally, while multi-objective models like RMTL and PLE lack mechanisms for explicit control, MocDT integrates such functionality, enabling precise adjustments to achieve desired outcomes across multiple objectives. This ability to control and optimize various targets concurrently makes MocDT superior to other models, including PHN and traditional DT models, by ensuring better performance and adaptability in multi-objective environments.

\begin{figure*}[!t]
\centering
\includegraphics[width=1\linewidth]{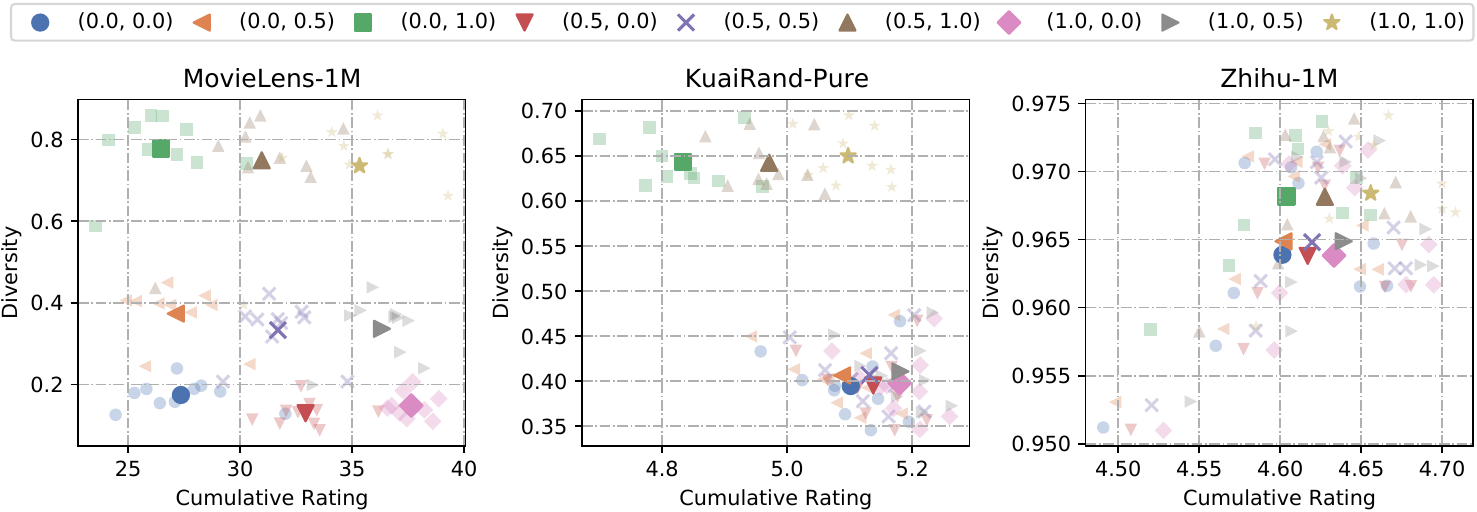}
\caption{Evaluation results conditioned on nine prioritized objectives. Here, the notation (1.0, 0.0) corresponds to scenarios where $O_{\text{rating}} = 1$ and $O_{\text{div}} = 0$, indicating a focus on rating maximization with no emphasis on diversity.}
\label{fig:main_result}
\end{figure*}

\begin{figure}[!t]
\centering
\includegraphics[width=1\linewidth]{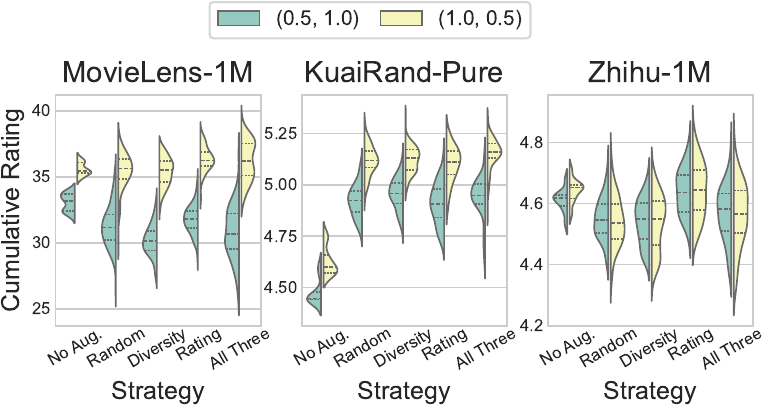}
\caption{Effect of augment strategies}
\label{fig:strategy}
\end{figure}

\subsection{Results on Controlling Ability (RQ2)}
\label{sec:exp-control}


\myfig{main_result} showcases the influence of predefined objective points on the evaluation metrics within our MocDT model framework. These objective points are represented as tuples, with the first and second elements corresponding to the cumulative rating and diversity, respectively. Upon the stabilization of our model, we subjected it to tests against a set of nine characteristic objective points.
The testing protocol involved the generation of 10 items conditioned on a given objective point and the subsequent assessment of Rating and Diversity metrics post each training epoch. Our methodology encompassed a total of 30 training epochs, with the outcomes from the concluding 10 epochs rendered as lighter patterns in the figure to depict transient states, while the darker symbols articulate the averaged results, offering a robust measure of performance consistency. 
The empirical results underscore the pivotal role of the preset objective point in directing the recommendation system's performance. Notably, the relative positioning of outcomes corresponding to the nine objective points remained invariant throughout the epochs, implying a reliable multi-objective control capability inherent to the MocDT model.

The visualization in \myfig{main_result} reveals a clear clustering effect within the MovieLens dataset for the nine objective points, highlighting the model’s ability to effectively learn the mapping from multiple objectives to recommended items. This clustering underscores a strong collaborative filtering signal, with each objective point forming distinct clusters.

In contrast, models trained on the KuaiRand and Zhihu datasets display some fluctuations across epochs, yet maintain stable relative positioning for each objective point, indicating robust multi-objective control. For KuaiRand, there is an inverse relationship between Diversity and Cumulative Rating. As the rating emphasis increases, diversity decreases, suggesting that in contexts like short videos where user preferences are homogeneous, overly diverse recommendations may reduce engagement.

Conversely, the Zhihu-1M dataset shows a positive correlation between Diversity and Cumulative Rating, with both increasing together. This indicates that in Zhihu’s content-rich environment, enhancing diversity aligns with user preferences, potentially boosting user satisfaction.

\subsection{Evaluation on Data Augmentation (RQ3)}
To investigate how data augmentation affects performance,  we explored variations in both the augmentation strategy and augmentation rate. The findings are depicted in \myfig{strategy} and \myfig{rate}, where, for clarity in observation, we chose to report only one metric: the Cumulative Rating. 
We employed two distinct objective points, (0.5, 1.0) and (1.0, 0.5), to show the effects with different prioritized objectives.

\subsubsection{Effect on Augment Strategies.} 
The analysis presented in \myfig{strategy} indicates distinct responses to augmentation strategies across the three datasets. In the KuaiRand-Pure dataset, augmented data markedly improved performance over the non-augmented set. This can be ascribed to the inverse relationship between Diversity and Cumulative Rating identified in \mysec{exp-control}. 
Augmentation introduces a breadth of data, enabling the model to better learn the mappings from multiple objectives to actions. All four augmentation strategies yield similar enhancements, diversifying the initially uniform dataset.

In contrast, MovieLens-1M and ZhihuRec-1M datasets exhibit negligible changes in Cumulative Rating with data augmentation, attributable to their already diverse nature encompassing varied scenarios. However, the Rating-focused augmentation strategy outperforms the random and Diversity augmentations, resonating with its intent to foster high-rating user-item interactions, thus refining the model's proficiency in the Rating metric.



\subsubsection{Effect on Augmentation Rates.}
\begin{figure}[!t]
    \centering
    \includegraphics[width=1\linewidth]{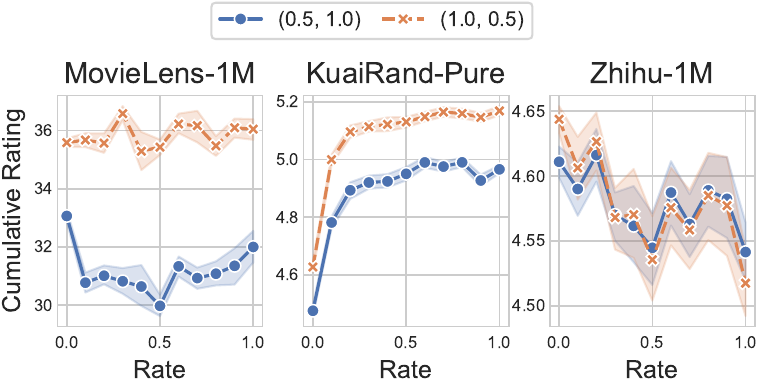}
    \caption{Effect of augment rates}
    \label{fig:rate}
\end{figure}

\myfig{rate} illustrates that increasing augmentation rates do not significantly affect the MovieLens dataset's Cumulative Rating, likely due to its intrinsic diversity and adequate collaborative filtering signals, hinting that excess data does not translate into performance gains. Conversely, the KuaiRand dataset shows a steady rise in metrics with higher augmentation rates, in line with the positive orientation of data augmentation, assisting the model in navigating the complexities of the negatively correlated indicators, thereby enhancing inference accuracy.

The Zhihu dataset, however, presents a contrasting pattern; as augmentation volume grows, the metric decreases. This may stem from the positive correlation between the evaluation metrics, where elevating one typically increases the other. Nonetheless, augmentation introduces a significant count of unconventional instances, acting as negative samples thereby distorting the training data distribution and ultimately hindering the model's ability to learn authentic user preference patterns reflected in the test set.

\subsection{Ablation Study (RQ4)}
\smallskip\noindent\textbf{Effect of Transformer layers.}
To investigate the effectiveness of the architecture in \myfig{framework}, we experiment with various combinations of step and sequence transformer structures and layer counts, finding fundamentally similar results. A symmetric approach of pairing one layer of Step Transformer with one layer of Sequence Transformer proved the most effective. We provide an ablation study for different layers for the objective point (1.0,1.0) on Movielens. 

The experimental results, as depicted in \mytable{layer}, demonstrate that the number of Transformer layers has a minimal impact on the Cumulative Rating, which fluctuates within a relatively narrow range. However, there is a notable effect on the Diversity of the recommended list. Specifically, a larger number of layers tend to yield higher Diversity in the recommendations. This suggests that the adaptability of the recommendation system improves as the number of Transformer layers increases, enabling the system to generate more diverse recommendation outputs.

\begin{table}[!t]
\renewcommand\arraystretch{1.2}
\caption{Effect of Transformer layers}
\label{tab:layer}
\begin{tabular}{@{}lccccc@{}}
\toprule
\textbf{\# Layer}     & \textbf{1} & \textbf{2} & \textbf{3} & \textbf{4} & \textbf{5} \\ \midrule
\textbf{Rating}    & 40.304     & 36.825     & 39.811     & 41.158     & 39.389     \\
\textbf{Diversity} & 0.543      & 0.573      & 0.690      & 0.808      & 0.730      \\ \bottomrule
\end{tabular}
\end{table}

\smallskip\noindent\textbf{Effect of future sequence length $H$.}
Computation of objectives is conducted on the $H$-step sequence, as formalized in \myeq{def_rating} and (\ref{eq:def_diversity}). We report the performance for the objective point (1.0, 1.0) on the Movielens dataset for different $H$ values in \mytable{window}.

The results show that smaller $H$ values yield better performance due to less divergence in future trajectory predictions, making them easier to forecast. However, very small $H$ values are impractical as we are typically interested in the long-term benefits of a strategy, thus we cannot solely pursue performance at low $H$ values. Hence, we chose $H=10$ for other experiments in this study.

\begin{table}[!t]
\caption{Effect of the length of the future sequence}
\label{tab:window}
\begin{tabular}{@{}llllll@{}}
\toprule
\textbf{H}         & \textbf{3} & \textbf{5} & \textbf{8} & \textbf{10} & \textbf{15} \\ \midrule
\textbf{Rating}    & 12.463     & 20.677     & 32.409     & 40.468      & 56.046      \\
\textbf{Rating/H}  & 4.154      & 4.135      & 4.051      & 4.047       & 3.736       \\
\textbf{Diversity} & 0.910      & 0.927      & 0.778      & 0.778       & 0.786       \\ \bottomrule
\end{tabular}
\end{table}

\subsection{Case Study}
To facilitate understanding of our recommendation results, we present a case study using the Movielens dataset for user ID=6040. After training the model for 30 epochs, we generate two recommendation lists, each with a length of $10$, corresponding to the objective points $(1.0, 0.0)$ and $(0.0, 1.0)$, respectively.

The results, as presented in \mytable{case}, detail the ratings and categories for each item in the recommendation list. For the objective point $(1.0, 0.0)$, the top 10 recommended movies achieved a cumulative rating of $42.812$ with a Diversity of $0$, indicating that all movies belonged to the same category. Conversely, for the objective point $(0.0, 1.0)$, the top 10 recommended movies were diverse across categories, yielding a Rating of $21.607$ and a Diversity of $0.902$. This case study highlights the impact of prioritizing different objectives and demonstrates the adaptability of MocDT to various objectives during the inference stage.

\begin{table}[!t]
\caption{Case Study with Two Extreme Objectives: The objective (1.0, 0.0) focuses on maximizing cumulative rating while minimizing diversity, whereas (0.0, 1.0) prioritizes diversity at the expense of cumulative rating.}
\label{tab:case}
\tabcolsep=1.5pt
\renewcommand\arraystretch{1.3}
\begin{tabular}{lcccccccccc}
\Xhline{1.5pt}
\rowcolor{mygray}
\multicolumn{11}{c}{\normalsize{\textbf{For objective point (1.0, 0.0)}}}  \\ 
\hline
\textbf{Position} & 1                                                    & 2                                            & 3 & 4                                            & 5 & 6                                              & 7 & 8 & 9 & 10 \\ \hline
\textbf{Rating}   & 4.61                                                          & 4.13                                                  & 4.33       & 4.20                                                  & 4.40       & 4.21                                                    & 4.13       & 4.12       & 4.04       & 4.66        \\
\textbf{Category} & {[}8{]}                                                       & {[}8{]}                                               & {[}8{]}    & {[}8{]}                                               & {[}8{]}    & {[}8{]}                                                 & {[}8{]}    & {[}8{]}    & {[}8{]}    & {[}8{]}     \\ 
\Xhline{1.5pt}
\rowcolor{mygray}
\multicolumn{11}{c}{\normalsize{\textbf{For objective point (0.0, 1.0)}}}  \\ 
\hline
\textbf{Position} & 1                                                    & 2                                            & 3 & 4                                            & 5 & 6                                              & 7 & 8 & 9 & 10 \\ \hline
\textbf{Rating}            & 1.71                                                          & 2.42                                                  & 2.03       & 1.26                                                  & 3.05       & 1.77                                                    & 1.68       & 3.30       & 3.05       & 1.37        \\
\textbf{Category}          & $\begin{bmatrix} 1 \\ 15 \\ 16 \end{bmatrix}$ & $\begin{bmatrix} 3 \\ 4 \end{bmatrix}$ & {[}5{]}    & $\begin{bmatrix} 4 \\ 5 \end{bmatrix}$ & {[}5{]}    & $\begin{bmatrix} 11 \\ 15 \end{bmatrix}$ & {[}16{]}   & {[}8{]}    & {[}8{]}    & {[}11{]}   \\
\Xhline{1.5pt}
\end{tabular}
\end{table}

\section{Related Work}
\label{sec:related}

In this section, we briefly introduce the multi-objective recommendation task, the concept of controllable recommendation, and the application of Decision Transformers in recommender systems.

\subsection{Multi-Objective Recommendation}
Multi-objective recommendation systems, also referred to as multi-task recommendation frameworks, aim to simultaneously optimize multiple recommendation objectives \citep{jannach2023survey}. Unlike traditional systems that mainly focus on predicting the relevance of items to users, multi-objective recommenders embrace a wider array of goals that more accurately reflect the complexity of real-world scenarios. These goals span various domains, which can be categorized into quality objectives, multistakeholder objectives, and so on \cite{JannachMORSsurvey}. Our research primarily investigates various quality objectives, or in other words, diverse recommendation performance metrics.

The complexity of multi-objective recommendation systems stems from the fact that their goals are often correlated or even competing. 
In addressing these challenges, researchers have developed various methodologies. 
A prevalent approach involves re-ranking item lists according to different objectives \citep{JannachMORSsurvey}. For example, \citet{LiRankEnsemble} studies how to fuse multiple item rankings by learning personalized item-level ensemble weights. Others have leveraged shared-weight network architectures to manage multiple objectives \citep{Tang2020ProgressiveLE, Ma2018ModelingTR}.
Further aligning with our research are explorations into the use of RL to resolve this challenge. For example, \citet{CaiConstrainedAC} introduces auxiliary objectives as constraints within the policy optimization process, and \citet{LiuRLMultiTaskRec} incorporates diverse recommendation tasks within a session-based framework, employing RL to dynamically adjust weights assigned to each objective's loss function.


\subsection{Multi-Objective Controllable Recommender}
With the rapid advancement of cutting-edge models, particularly large language models (LLMs) like ChatGPT, their extensive application in recommender systems \cite{surveyLLM4rec} has shifted the paradigm of AI-driven services from traditional Software-as-a-Service (SaaS) \cite{SaaS} to Model-as-a-Service (MaaS) \cite{MaaS}. This transition enables users to interact directly with AI models, significantly increasing the demand for model controllability. 
As highlighted in the survey by \citet{shen2024survey}, controllable learning in information retrieval encompasses various definitions and applications, including multi-objective control. Under this framework, the model must adapt to user preferences for each objective, which can be expressed explicitly (e.g., through weights or natural language) or implicitly (e.g., via interaction behaviors).

In multi-objective optimization, or Pareto optimization, the goal is to optimize multiple objectives simultaneously. A common approach is scalarization \citep{Ozan2018MTLasMO}, which transforms the problem into a single-objective one by constructing a weighted sum of losses for each objective. This enables the optimization of diverse recommendation objectives, as shown in \citep{lin2019pareto, Xie2021Pareto, Pareto22wsdm}. The weights form a preference vector, allowing controlled tuning of the model's optimization path. Notably, \citet{Navon2020LearningTP} introduced a framework using Pareto Hypernetworks to encode preference vectors and generate aligned outcomes. Extending this, \citet{Chen2023Control} proposed CMR, which integrates Pareto Hypernetworks to produce model parameters tailored to specified objectives.
However, CMR differs significantly from our approach. It relies on reinforcement learning and a pre-trained reward model, making performance dependent on reward model design. Its computational complexity is at least $O(M^2)$ per item, where $M$ is the number of candidates, limiting it to re-ranking tasks and reducing its applicability to open datasets.
Additionally, \citet{shen2024generating} propose using a diffusion model to generate recommender parameters conditioned on objectives. However, training targets based on backbone model parameters may be suboptimal due to potential instability. Meanwhile, \citet{lu-etal-2024-aligning} leverage LLMs for category control, but their performance depends on semantic understanding and collaborative filtering, risking biases \cite{gao2024sprec}.

These advancements lay the groundwork for controllable optimization in multi-objective recommendation systems.

\subsection{Decision Transformer for Recommendation }

The Decision Transformer (DT), introduced by \citet{NEURIPS2021_7f489f64}, reimagines offline reinforcement learning (RL) as a sequence modeling problem. With the growing adoption of RL techniques in recommendation systems \cite{easyRL4Rec,wentao2023billp}, DT has emerged as a promising alternative to conventional offline RL algorithms. Recent studies have explored and validated the potential of DT in the recommendation domain \cite{li2024decision,gunal-etal-2024-conversational,lee2023clinical}.
For example, \citet{Wang2023CausalDT} extended DT's application by predicting an action's immediate reward based on its causal effects. Similarly, \citet{10.1145/3543507.3583418} focused on enhancing user retention by integrating a novel reward embedding aggregation mechanism and employing weighted contrastive learning to improve DT's performance. However, both approaches are limited to single-objective scenarios, restricting their applicability to real-world recommendation tasks that typically involve multiple, often conflicting, objectives.
Some works have attempted to integrate both short-term and long-term rewards into DT models \cite{liu2024sigir} or optimized multiple objectives by considering inter-session and intra-session returns simultaneously \cite{10.1145/3627673.3679533}. However, these approaches do not account for dynamically changing objectives during the inference stage. 

To date, no research has adapted DT for industrial recommendation environments, where policies must quickly adjust to dynamic combinations of multiple objectives in real time.

\section{Conclusion}



This study addresses a critical challenge in recommender systems: the dynamic adjustment of multiple objectives during the inference stage. Traditional methods require retraining, which is highly inflexible. To overcome this limitation, we propose the Multi-Objective Controllable Decision Transformer (MocDT), which introduces a control signal into the Decision Transformer architecture, enabling the model to generate item sequences aligned with predefined objectives without the need for retraining.
Empirical results demonstrate MocDT's effectiveness in controlling and adapting to multi-objective item sequence generation. Furthermore, our research sheds light on the impact of data augmentation on model learning across various settings, offering valuable insights for optimizing multi-objective recommendations in diverse data contexts.

While our work marks significant progress, it acknowledges current limitations, including the inability to fully verify the Pareto optimality of specified objectives and the feasibility of objectives provided to the model. Future research will explore the integration of controllable multi-objective generation in generative models and large language models to further enhance the personalization and adaptability of recommendation systems.




\bibliographystyle{ACM-Reference-Format}
\bibliography{MocDT}

\end{document}